\documentclass{Interspeech2024}
\pdfoutput=1
\interspeechcameraready

\usepackage{multirow,color} 
\usepackage{ulem}
\usepackage{cite}
\usepackage{authblk}
\usepackage[marginal]{footmisc}

\title{Refining Self-Supervised Learnt \\
Speech Representation using Brain Activations\vspace{-1.2em}\thanks{* Corresponding author. This work is supported by the National Natural Science Foundation of China (62101523), Hefei Municipal Natural Science Foundation (2022012) and USTC Research Funds of the Double First-Class Initiative (YD2100002008).}}

\name {Hengyu}{Li}
\name {Kangdi}{Mei}
\name {Zhaoci}{Liu}
\name {Yang}{Ai}
\name {Liping}{Chen}
\name {Jie}{Zhang*} 
\name {Zhenhua}{Ling}

\address{National Engineering Research Center of Speech and Language Information Processing, \\
University of Science and Technology of China, Hefei, P.R.China}
\email{lihengyu@mail.ustc.edu.cn, \{jzhang6,zhling\}@ustc.edu.cn}

\keywords{Pre-trained speech model, wav2vec2.0, brain activation, SUPERB}

\begin{document}

\maketitle

\begin{abstract}
It was shown in literature that speech representations extracted by self-supervised pre-trained models exhibit similarities with brain activations of human for speech perception and fine-tuning speech representation models on downstream tasks can further improve the similarity. 
However, it still remains unclear if this similarity can be used to optimize the pre-trained speech models. 
In this work, we therefore propose to use the brain activations recorded by fMRI to refine the often-used wav2vec2.0 model by aligning model representations toward human neural responses. 
Experimental results on SUPERB reveal that this operation is beneficial for several downstream tasks, e.g., speaker verification, automatic speech recognition, intent classification. One can then consider the proposed method as a new alternative to improve self-supervised speech models.
\end{abstract}


\section{Introduction}

Self-supervised learning has revolutionized the field of speech processing, where the resulting model can extract some efficient, robust and universal features from unlabeled samples through e.g., adversarial learning, clustering~\cite{jaiswal2020survey}. Given a small amount of labeled data, the model can be fine-tuned to further upgrade the universal features with representation capabilities into exclusive features for downstream tasks \cite{liu2021self}. In \cite{DBLP:journals/corr/abs-1807-03748}, contrastive predictive coding (CPC) was proposed,  which leverages adversarial learning to learn audio features. Schneider et al. \cite{schneider2019wav2vec} followed this scheme and optimization objectives of CPC and found that using the pre-trained wav2vec model to extract features can effectively improve the performance of automatic speech recognition (ASR). Baevski et al. \cite{baevski2019vq} further proposed a quantization-based wav2vec, followed by  a more advanced end-to-end self-supervised model, called wav2vec2.0~\cite{baevski2020wav2vec}, which is shown to be more promising for ASR. HuBERT was then proposed in~\cite{hsu2021HuBERT}, which provides a discretized target sequence containing acoustic and semantic information, enabling to learn richer multi-level speech features  and resulting in a stronger generalization to various downstream speech processing tasks.


Due to the impressive efficacy of self-supervised speech pre-tained models~\cite{mehrish2023review}, numerous studies \cite{krishnan2022self,zhang2023basen,zhang2023sparsity} make efforts to establish connections between the general features extracted from these models and the information processing procedures of the human brain's nervous system~\cite{kanwisher2023using}. For example, Millet et al. \cite{millet2022toward} conducted an alignment-matching study using wav2vec2.0 \cite{baevski2020wav2vec} and functional magnetic resonance imaging (fMRI) brain activity records during speech listening. The result shows a correlation between the speech representation of pre-trained models and brain activations during speech perception. Similarly, Vaidya et al. \cite{vaidya2022self} validated that self-supervised speech models (e.g., APC \cite{chung2020vector}, wav2vec, wav2vec2.0, HuBERT) can effectively capture information levels that are related to different stages of human cortical speech processing. Oota et al. \cite{oota2023neural} then evaluated the performance of different categories of speech models in encoding speech stimuli toward human neural activations. It was further indicated in~\cite{oota2023speech} that using downstream tasks to fine-tune pre-trained models can improve the neural encoding performance.


Conversely, it is still unclear whether leveraging neural activations in the human brain can enhance the performance of pre-trained speech models on downstream tasks. For text processing, Toneva et al. \cite{toneva2019interpreting} suggested that updating a pre-trained language model to better predict the neural responses of human language processing can improve language understanding capacity.  Schwartz et al. \cite{schwartz2019inducing} demonstrated that using brain activity records to fine-tune BERT \cite{kenton2019bert} can enhance the ability to predict brain neural activity during language processing without compromising its performance on downstream tasks. However, there is currently no report on  how to  optimize the pre-trained speech models via neural encoding for downstream speech tasks. Hence, it is worthwhile to investigate similar approaches for enhancing self-supervised pre-trained speech models \cite{tuckute2022many}, which are thus called \textbf{neuroscience-driven}.


Inspired by \cite{oota2023speech}, in this work we make efforts to explore the potential  of pre-trained speech models on downstream tasks by incorporating neural activations from human speech perception into model parameters. Specifically, the well-known wav2vec2.0 is selected as the examplary pre-trained speech model without loss of generality. The proposed method refines wav2vec2.0 with brain activations by adding convolutional and linear layers on top of the model. These layers aim to predict brain activations from speech signals, and the parameters of wav2vec2.0 are updated using the L2-regularized mean square error (MSE) loss. Additionally, historical audio information, based on the predictive coding theory \cite{caucheteux2023evidence}, is utilized as the  input for prediction. After fine-tuning, the refined model is evaluated using the general speech processing performance benchmark (SUPERB) \cite{yang2021superb}. The obtained results verify the efficacy of the proposed method on several downstream tasks.
\begin{figure}[t]
  \centering
  \includegraphics[width=7.5cm]{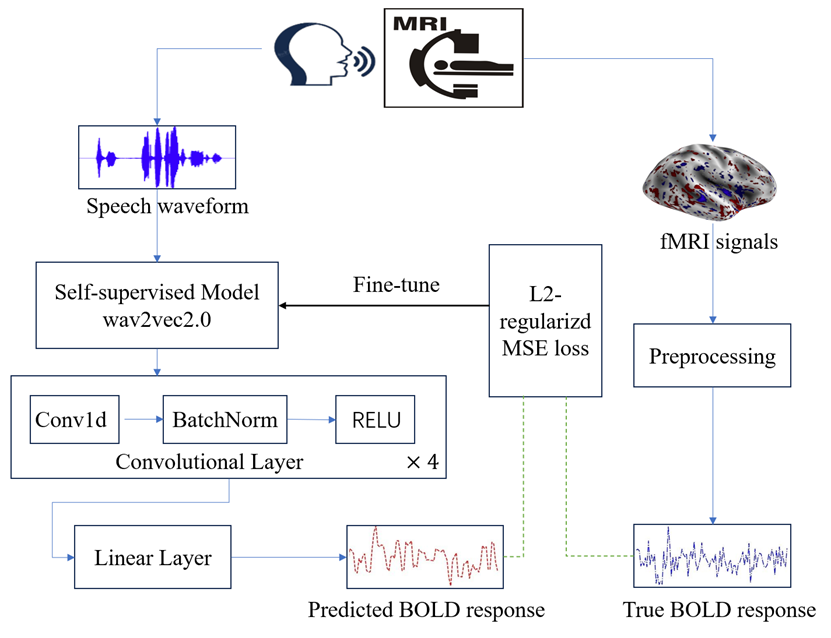}
  \caption{Flowchart of refining wav2vec2.0 using brain signals.}
  \label{modelStructure}
  \vspace{-0.5cm}
\end{figure}

\section{Methodology}
\subsection{Material and Pre-processing}
\label{ssec:subhead}

The “tunnel under the world” subset of the “Narratives" dataset \cite{nastase2021narratives} is adopted throughout  experiments in this paper, which comprises various fMRI data collected from human subjects when listening to natural oral stories.

The audio stimulus includes 3435 words at a length of 1534 seconds. Silences of 3 seconds and 23 seconds are inserted before and after the stimulus, respectively, resulting in a total fMRI scan duration of 1560 seconds. This scan consists of 1040 repetition times (TR) with 1.5 seconds per TR. During each TR, the scan outputs blood oxygen level dependent (BOLD) signals, which are widely-used for analyzing brain activation and functional connectivity. These signals are measured based on the magnetic changes between the oxygenated and deoxygenated states of hemoglobin in the blood. The fMRI data used in this study originate from the dataset published on datalad\footnote{\url{http://datasets.datalad.org/?dir=/labs/hasson/narratives}} \cite{nastase2021narratives}.

We exclude subjects 004 and 013 as recommended in the dataset. The neural responses of the remaining 21 subjects are utilized in experiments. These fMRI data have to be pre-processed before being applied to refine the pre-trained speech model. Initially, the brain activations represented by BOLD signals at each TR are projected onto the surface space of `fsaverage6' \cite{fischl2012freesurfer}, composed of 40962 voxels. Subsequently, only the voxels corresponding to the brain regions of interest (ROIs) related to auditory and language processing are selected. The Glasser atlas \cite{glasser2016multi} is adopted for this purpose, partitioning each hemisphere into 180 ROIs. The selected voxels are from ROIs of the early auditory cortex (EAC), auditory association cortex (AAC) and inferior frontal gyrus (IFG). Finally, the pre-processed BOLD signals in each TR are represented by a vector of 5085 dimensions, corresponding to 2468 voxels in the right brain and 2617 voxels in the left brain.


\subsection{Refining wav2vec2.0 by Predicting Brain Activations}
\label{ssec:subhead}
\subsubsection{Problem Description}
\label{sssec:subsubhead}

The goal of this work is to construct a neural encoding model by adding additional layers on top of the pre-trained wav2vec2.0, which can predict  subject's brain activations given the speech stimulus. Due to the data conditions in Section 2.1, the model output is the 5085-dimensional pre-processed BOLD vector at each TR, and the model input comprises 16kHz speech waveforms corresponding to the current and previous TRs. The wav2vec2.0 model is refined by optimizing the prediction error. It is thus interesting to investigate whether this refinement can enhance the performance of wav2vec2.0 on downstream tasks. 
\subsubsection{Vanilla wav2vec2.0 Model}
\label{ssec:subsubhead}

In this work, as an example Wav2vec2.0 \cite{baevski2020wav2vec} is adopted as the self-supervised speech model. It integrates the Gumbel softmax quantization module from vq-wav2vec \cite{baevski2019vq} and BERT \cite{kenton2019bert} in an end-to-end fashion. 
In wav2vec2.0, raw speech waveforms first pass through a series of convolutional layers to generate latent speech representations $\textbf{Z}$, which are then masked and fed into the Transformer to obtain context representations $\textbf{C}$. Simultaneously, quantized representations $\textbf{Q}$ are obtained via product quantization on $\textbf{Z}$. The self-supervised loss function for pre-training comprises a contrastive loss and a diversity loss. The former discriminates positive samples from negative ones for the quantized representations at the masked positions. Meanwhile, the latter loss aims to enhance the representation ability of the quantization codebook by encouraging the model to utilize all codebook entries equally. After pre-training, the model can be applied to downstream tasks by cascading task-specific layers and fine-tuning the entire model using labeled data in combination with task-specific loss function(s).

\begin{table*}[t]
\centering
\caption {The results of vanilla and refined wav2vec2.0 models on  SUPERB. 
For ASR,  the WER is evaluated without language models. The result better than vanilla-base for each task is highlighted in bold. The stars indicates a significant improvement ($p<0.05$) in one-tailed paired t-test between refined and vanilla models.}
\vspace{-0.2cm}		
\label{table1}
\begin{tabular}{|c|c|c|c|c|c|c|c|c|c|c|c|c|}
\hline
\multicolumn{2}{|c|}{\multirow{2}{*}{Model}} &\multicolumn{1}{c|}{PR}&\multicolumn{1}{c|}{KS}&\multicolumn{1}{c|}{IC}&\multicolumn{1}{c|}{SID}&\multicolumn{1}{c|}{ER}&\multicolumn{1}{c|}{ASR}&\multicolumn{2}{c|}{SF}&\multicolumn{1}{c|}{ASV}&\multicolumn{1}{c|}{Overall}\\
\cline{3-12}

\multicolumn{2}{|c|}{} & PER $\downarrow$ & ACC $\uparrow$ & ACC $\uparrow$ & ACC $\uparrow$ & ACC $\uparrow$ &WER $\downarrow$ & F1 $\uparrow$ & CER $\downarrow$ & EER $\downarrow$  & ${\rm superb}_s$ $\uparrow$\\
\hline

\multicolumn{2}{|c|}{Vanilla-base} &6.03  &96.23 &92.57	&74.45	&62.53	&6.60	&\textbf{87.65}	&25.83	&5.98 &0\\ 
\hline

\multicolumn{2}{|c|}{Vanilla-large(LV-60)} &4.75	&96.66	&95.28 &86.14	&65.64	&3.75	&87.11	&27.31	&5.65  &1000 \\ 
\hline

\multicolumn{2}{|c|}{Stimuli-pretrain} &6.22	&95.72	&91.93	&73.88	&62.23	&6.70	&86.25	&27.07	&6.08 &-293.43\\ 
\hline

\multicolumn{2}{|c|}{Refined} &\textbf{5.67*}	&96.23	&\textbf{93.78*}	&\textbf{75.28*}	&\textbf{63.72}	&\textbf{6.36*}	&87.31	&\textbf{25.15}	&\textbf{5.50*}  &\textbf{388.59}\\ 
\hline

\end{tabular}
\label{table_downstreamtask}
\vspace{-0.1cm}
\end{table*}

\subsubsection{Model Architecture}
\label{sssec:subsubhead}

The structure of the neural encoding model is shown in Fig. 1. For each TR, the speech waveforms of the current TR and the previous $n-1$ TRs are utilized as the model input. The reason of using historical speech waveforms is twofold: 1) the asynchronous relationship between the speech stimulus and BOLD signals is considered, in line with the mechanism of BOLD acquisition. Due to the inertia of changes in blood oxygen levels, the BOLD response reaches its peak after a latency period of several seconds, making historical information relevant. 2) the predictive coding theory \cite{caucheteux2023evidence} suggests that the human brain can effectively utilize long-term information in speech. Therefore, incorporating historical information into the speech input may enhance the model's ability to perform neural encoding.


In Fig.1, the input speech waveforms of $1.5n$ seconds are initially processed by wav2vec2.0, and the computed context representations $\textbf{C}$ are employed to predict the BOLD vector at the current TR. In wav2vec2.0, the frame rate of context representations is 50Hz, and the input speech waveform consist of $75n$ frames in total. To downsample the context representations, four convolutional layers with strides of $\{n, 5, 5, 3\}$ are inserted. A z-score standardization \cite{schober2021statistics} is then applied to the output of the convolution layers. Following the standard neural encoding models \cite{caucheteux2021model}, a linear layer is added at the end of the model to predict the BOLD response at the current TR.



An L2-regularized MSE loss is used to train the neural encoding model. Considering the limited amount of fMRI data, a two-stage strategy is utilized. After pre-training wav2vec2.0 and randomly initializing the convolutional and linear output layers, the first stage only focuses on updating the linear layer while freezing other modules. Once the training of this stage converges, the linear layer is frozen, and the rest of the model, including wav2vec2.0, is updated at the second stage.

\section{Experiments and Results}
\label{sec:pagestyle}
\subsection{Implementation}
\label{ssec:subhead}
The vanilla-base wav2vec2.0 is provided by fairseq, which was already pre-trained on Librispeech-960h dataset\footnote{\url{https://github.com/facebookresearch/fairseq/tree/main/examples/wav2vec}\label{w2v}}. The vanilla-large(LV-60) wav2vec 2.0 Large is the comparison model originated from fairseq, featuring a larger network and pre-trained on LibriLight-60kh dataset\textsuperscript{\ref {w2v}}. For the proposed method, the TRs are shuffled and divided into a training set, a validation set and a test set at a ratio of 8:1:1. The validation set is employed to tune the L2 regularization weight $\lambda$, which varies between 1e-3 and 1e-1.
The four convolutional layers have strides of $\{n,5,5,3\}$  with a kernel size of 3, and the padding size of the first layer is 1.  Each layer performs batch normalization and the activation function is ReLU. 
At the first stage of the model training, 60 epochs are conducted using a basic learning rate of 3e-3, but the second stage uses 60 epochs at a learning rate of 3e-4. In the first 10\% epochs of both stages, the learning rate linearly increases from 0 to the  basic learning rate, and in the subsequent epochs it linearly decreases to 0. 


To observe the benefit of the alignment from additional training steps, we introduce a “stimuli-pretrain” model, which undergoes follow-up pre-training using the audio stimulus in the “Narratives” dataset with the same number of training steps as the refined training for comparison. The training is performed upon the same configuration file as in the vanilla-base model.
All the trainings are conducted on a server equipped with 4 Nvidia A100 GPUs, each with 80GB GPU RAM.

In theory, the BOLD signals induced by the stimulus gradually rise after about 1-2s and reach the peak at 5-6s\cite{dale1997selective}.  This characteristic is roughly consistent in the motor, visual, and auditory fields~\cite{friston1998event}, but the change in prefrontal cortex (which is related to cognition and emotion) is slower than that in the visual area by 4s~\cite{schacter1997late}. Considering the differences in the prefrontal lobe, a relatively high level of BOLD signal will be achieved at a time offset of 9s, that is, $n$ = 6. In other words, setting the input history waveforms to 9s may enable the model to learn brain activation information more effectively. Therefore, we will use $n$ = 6 as an example in the subsequent analyses.

\subsection{Evaluation Tasks for Pre-trained Models}
\label{ssec:subhead}
SUPERB \cite{yang2021superb} aims to directly use pre-trained speech models on various downstream tasks by establishing a framework that uses pre-trained models with frozen parameters and lightweight prediction heads tuned for each task. This framework reflects four categories of speech tasks: content, speaker, semantics and paralinguistics. We evaluate the performance on 3 content tasks including phoneme recognition (PR), ASR and keyword spotting (KS), 2 speaker tasks of speaker identification (SID) and automatic speaker verification (ASV), 2 semantic tasks including intent classification (IC) and slot filling (SF), and 1 paralinguistics task of emotional recognition (ER). When designing and training these prediction heads for downstream tasks, we follow the default configuration of SUPERB, except that the learning rate of  1e-3 is used for the SID task. The evaluation results of all compared models are obtained by applying them to downstream tasks following the SUPERB framework with our own implementation, except the results of vanilla-large which are published ones \cite{yang2021superb}. 

In SUPERB, superb-score (${\rm superb}_s$) is utilized to measure the overall performance of upstream models\footnote{\url{https://superbbenchmark.org/challenge-slt2022/metrics}}.
For each downstream task $t$, we define a specific metric ranging from [0, 1000] to evaluate the performance of the refined model $u$, following the definition of ${\rm superb}_s$. By weighting the metrics on all the $T$ tasks, the overall performance can then be calculated, where the vanilla-large (LV-60) is used as the upper limit of the evaluation and vanilla-base as the lower bound, given by
\begin{equation}
{\rm superb}_s=\frac{1}{|T|}\sum_{t\in \{1,\cdots,T\}} \frac{1000\left(s_t (u)-s_t (\text{base})\right)}{s_t (\text{large})-s_t (\text{base})}.
\end{equation}
Since the upper limit is worse than the lower bound on the SF task~\cite{yang2021superb}, the metric ${\rm superb}_s$ is excluded for the SF task.


\subsection{Experimental Results}
\label{ssec:subhead}
First, we use a fixed $n$ to study the impact of refinement on downstream tasks. The results of the refined and vanilla models are shown in Table 1. We assess the reliability of our results by applying one-tailed paired t-test on each audio in the SUPERB test set for each downstream task. It shows that using an appropriate amount of historical audio information in the neural encoding model, the refined wav2vec2.0 model can achieve a similar performance with the vanilla on ER, KS and SF tasks, and outperform the vanilla  on  PR, IC, SID, ASR and ASV.

However, the downstream results are not significantly different from the vanilla wav2vec2 on some tasks. For the content tasks, we speculate that the proposed method is likely inhibited by the change in the audio domain on PR and ASR, both of which use the same domain as the vanilla-base. We thus evaluate the stimuli-pretrain model on same downstream tasks and the results are shown at the fourth row of Table 1. It can be found that the model will not benefit on downstream tasks from using additional data for pre-training, which can even degrade due to the domain shift. As the proposed refined model also experiences the domain migration through the alignment, the results of refining operations are thus promising for downstream tasks.

\begin{figure}[tb]
\vspace{-0.5cm}
\begin{minipage}[b]{1.0\linewidth}
  \centering
  \centerline{\includegraphics[width=7.5cm]{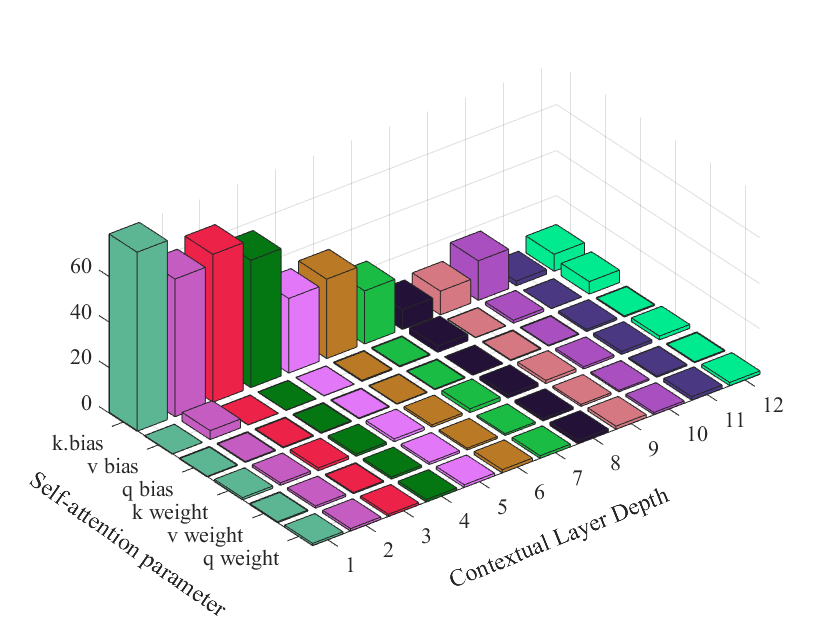}}
\end{minipage}
\caption{The percentages of parameter changes after refining different parameter types at different model layers. 
}
\label{fig:layer}
\vspace{-0.5cm}
\end{figure}

\begin{figure}[tb]
\begin{minipage}[b]{.40\linewidth}
  \centering
  \centerline{\includegraphics[width=4.2cm]{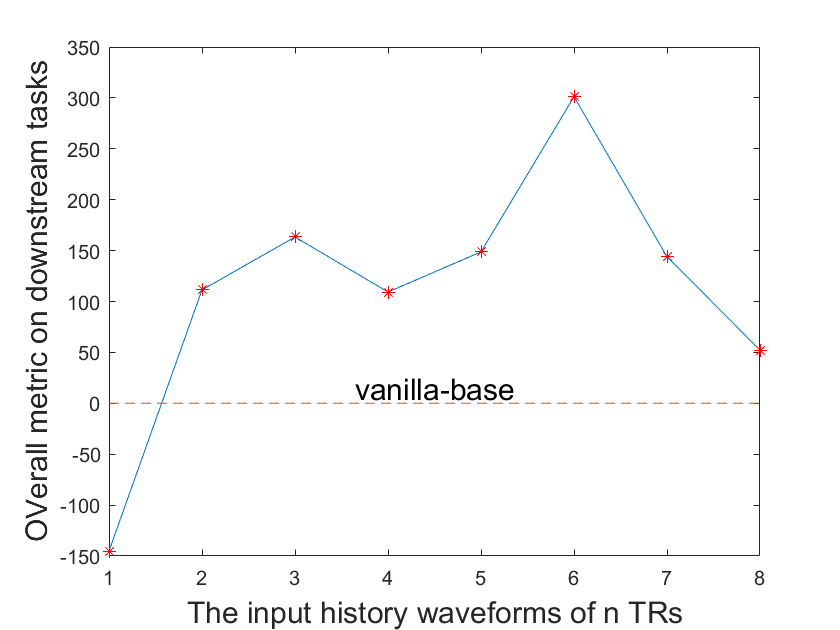}}
  \begin{center}{(a) }
  \end{center}
\end{minipage}
\hspace{1.0cm}
\begin{minipage}[b]{.40\linewidth}
  \centering
  \centerline{\includegraphics[width=4.2cm]{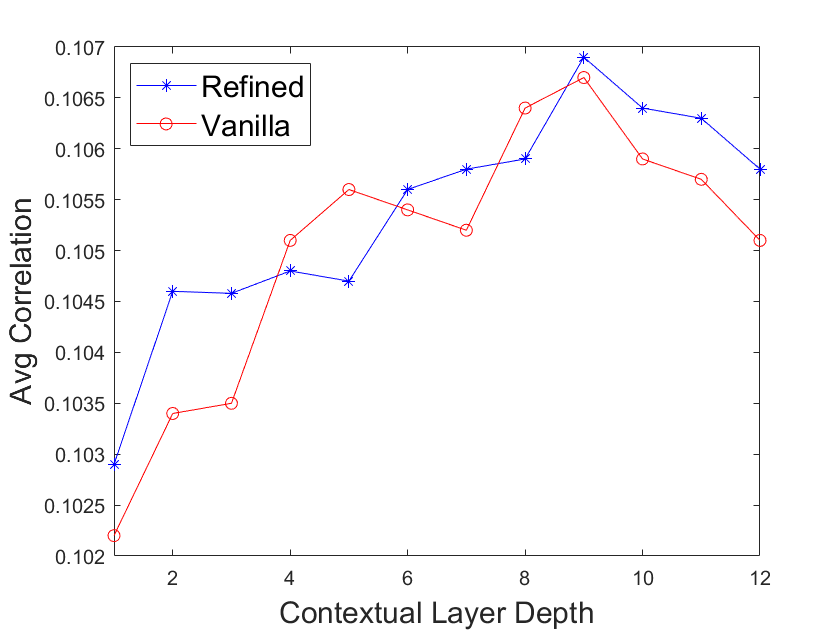}}
  \begin{center}{(b) }
  \end{center}
\end{minipage}
\caption{Analytical results of (a) the length of input history waveforms and (b) predicting brain activations.} 
\label{fig:layer}
\vspace{-0.5cm}
\end{figure}

\subsection{Experimental analysis}
\label{ssec:subhead}
\begin{figure}[tb]

\begin{minipage}[t]{.40\linewidth}
  \centering
  \centerline{\includegraphics[width=4.0cm]{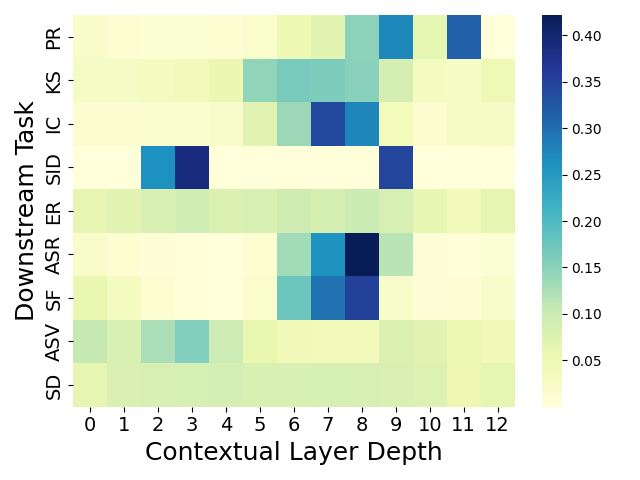}}
  \begin{center}{(a) Layer weights of the refined model on eight downstream tasks}
  \end{center}
\end{minipage}
\hspace{0.8cm}
\begin{minipage}[t]{0.40\linewidth}
  \centering
  \centerline{\includegraphics[width=4.0cm]{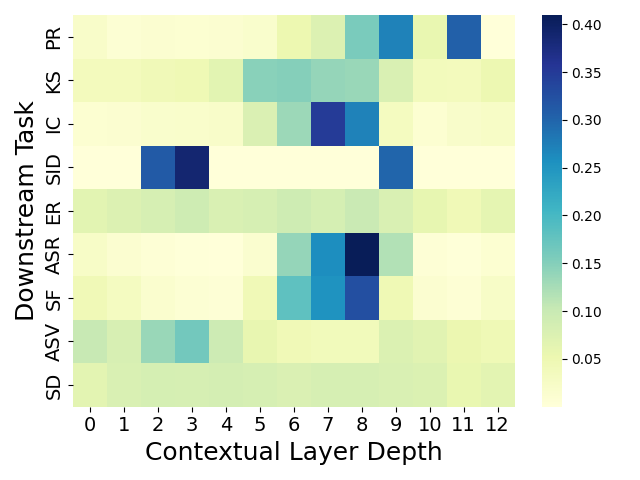}}
  \begin{center}{(b) Layer weights of the vanilla model on eight downstream tasks }\medskip
  \end{center}
\end{minipage}
\begin{minipage}[b]{1.0\linewidth}
  \centering
  \centerline{\includegraphics[height=4.2cm]{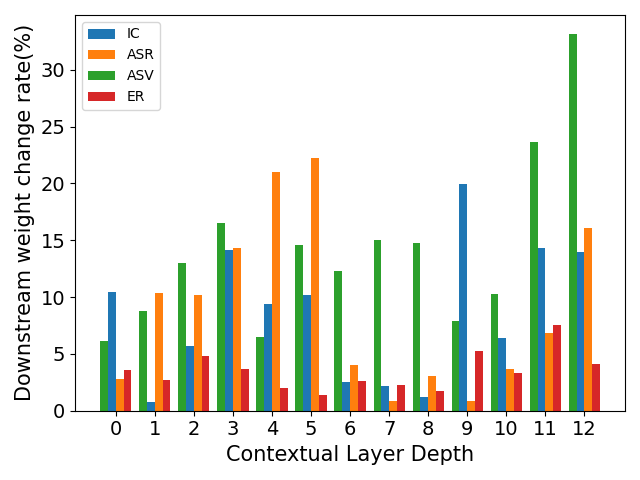}}
  \begin{center}
  {(c) Layer weights change rates on four representative downstream tasks}\medskip
  \end{center}
\end{minipage}
\hfill
\vspace{-0.5cm}
\caption{
Analytical results of layer weights.
}
\label{fig:res}
\vspace{-0.5cm}
\end{figure}
\label{ssec:subhead}

\textbf{Analysis of parameter changes after refining:}
Although we can conclude from aforementioned results that the refined model can extract better speech representations, an accurate illustration of how the speech representation is refined is not straightforward. For this,  we  examine how the parameters of  self-attention in the Transformer layers  change, which is shown in Fig.~2.
It can be observed that compared to vanilla, the bias of self-attention in the refined model has a significant change for $K$, while the biases of $Q$ and $V$ have relatively small changes. As the depth of the layer increases, the amount of changes in $K$ tends to decrease. This will lead the refined model to focus on different parts of the features, resulting in different attention distributions and positive effects on some downstream tasks.  Therefore, we speculate that such changes in attention distribution enable the model to learn relevant information on brain activation, thereby enabling a better understanding of speech.

\textbf{Analysis of the length of input history waveforms:}
In order to see the impact of the length of input history waveforms,  we adjust the value of $n$ from 1 to 8 and the result are shown in Fig.~3(a), where  $n$ = 1(-context) means that no history speech waveforms are considered for predicting the BOLD response at each TR.
Observing the average performance in terms of the context length $n$, we find that increasing the length of input history waveform first improves the performance of the refined model but then degrades. It does not follow a monotonic behaviour, which is probably caused by the fact that the acoustic and semantic understanding of human brain relies on both short-term and long-term cues contained in speech  \cite{caucheteux2023evidence}. This is consistent with the results in \cite{oota2023speech} that in the context of speech perception, human brains follow an information processing pipeline from word recognition, intention in sentences, emotion in stories, to speaker recognition.

\textbf{Analysis of predicting brain activations:}
To confirm the neural encoding ability of the refined model, we follow the method in \cite{millet2022toward} and compare the Pearson correlation coefficient (PCC) between the true brain activations and that predicted by the speech representation models before and after refining. Specifically, a ridge regression is adopted to predict the brain activations 
using the speech representations given by each layer of the vanilla or refined wav2vec2.0. Afterwards, the PCC is computed for each selected voxel, and the obtained results are shown in Fig.~3(b). It indicates that after refining the model, the PCCs between the predicted BOLD responses and true counterparts are generally higher. It is interesting that the maximum PCC is achieved in case of using the representations at the 9-th layer of wav2vec2.0 for prediction. This encourages to utilize such representations at middle layers to predict BOLD responses for model refinement in future works.

\textbf{Analysis of the change rates of layer weights after refining:}
Finally, as the SUPERB framework learns a weighted sum of the layer outputs for each downstream task, we carry out a deeper analysis on comparing the layer weights of refined and vanilla models on different downstream tasks, and calculate their change rates. The analytical results are presented in Fig.~4.
We observe that the change rates of layer weights are roughly inversely proportional to their absolute values, that is, the layer that plays an important role in a downstream task has a small change on its weight after being refined. The main function of our refined method is therefore to influence other secondary layers to better adapt to downstream tasks.

\section{Conclusion}
\label{sec:typestyle}
In this paper, we proposed a refinement-based wav2vec2.0 model by building a neural encoding model to predict brain activations using speech. It was validated that self-supervised speech pre-trained models can encode brain activation clues into model parameters and extract better speech representations, indicating feasibility of using the universal relation between speech representation and brain activation in self-supervised learning. It was also shown that using historical information to help predict the BOLD response at current TR and thus downstream tasks.
The proposed refining operation fills the gap between neuroscience and self-supervised  speech representation models. It should be insightful that more efforts can be made to optimize these models using multiple sources of auditory and linguistic information contained in the brain activities, particularly in low signal-to-noise ratio scenarios.


\normalem
\bibliographystyle{IEEEtran}
\bibliography{mybib}

\end{document}